\documentclass[aps, amsmath, preprint, amssymb,groupedaddress]{revtex4}
\setcitestyle{super}
\usepackage{graphicx}
\usepackage{gensymb}
\pdfoutput=1
\usepackage{amsmath}
\usepackage{color}
\usepackage{float}
\usepackage[english]{babel}
\usepackage{lipsum}
\usepackage{color,soul}

\newcommand{\beginsupplement}{%
        \setcounter{table}{0}
        \renewcommand{\thetable}{S\arabic{table}}%
        \setcounter{figure}{0}
        \renewcommand{\thefigure}{S\arabic{figure}}%
     }

\begin{document}
\title{Unexpected superconductivity at nanoscale junctions made on the topological crystalline insulator Pb$_{0.6}$Sn$_{0.4}$Te }

\author{Shekhar Das$^1$}
\author{Leena Aggarwal$^1$}

\author{Subhajit Roychowdhury$^2$}
\author{Mohammad Aslam$^1$}
\author{Sirshendu Gayen$^1$}
\author{Kanishka Biswas$^2$}
\email{kanishka@jncasr.ac.in}
\author{Goutam Sheet$^1$}
\email{goutam@iisermohali.ac.in}
\affiliation{$^1$Department of Physical Sciences,
Indian Institute of Science Education and Research Mohali,
Sector 81, S. A. S. Nagar, Manauli, PO: 140306, India}
\affiliation{$^2$New Chemistry Unit, Jawaharlal Nehru Center for Advanced Scientific Research, Jakkur, P.O., Bangalore 560064, India}

\begin{abstract}

Discovery of exotic phases of matter from the topologically non-trivial systems not only makes the research on topological materials more interesting but also enriches our understanding of the fascinating physics of such materials. Pb$_{0.6}$Sn$_{0.4}$Te was recently shown to be a topological crystalline insulator. Here we show that by forming a mesoscopic point-contact using a normal non-superconducting elemental metal on the surface of Pb$_{0.6}$Sn$_{0.4}$Te a novel superconducting phase is created locally in a confined region under the point-contact. This happens while the bulk of the sample remains to be non-superconducting and the superconducting phase emerges as a nano-droplet under the point-contact. The superconducting phase shows a high transition temperature $T_c$ that varies for different point-contacts and falls in a range between 3.7 K and 6.5 K. 
Therefore, this Letter presents the discovery of a new superconducting phase on the surface of a topological crystalline insulator and the discovery is expected to shed light on the mechanism of induced superconductivity in topologically non-trivial systems in general.

\end{abstract}

\maketitle

It is known that perturbations like magnetic dopant, structural distortion, mechanical strain and disorder etc. can be used to realize novel phases of matter out of the topological insulators (TIs) \cite{YS, Peng, Kirzhner, Satoshi, rmp1, rmp2} and the topological crystalline insulators (TCIs)\cite{sato, fu1, Fu, liu1, Zeljkovic, Tian, Mitrofanov,Ilija, okada, Erickson}. Based on these ideas, enormous amount of effort has been invested to create exotic superconducting phases from TIs\cite{Satoshi} and TCIs\cite{sato, fu1, balu} as the superconductors emerging from topologically non-trivial systems may show topological superconductivity and eventually may lead to the observation of Majorana modes in condensed matter systems\cite{Leijnse, majoran1, Been}. Among the potential candidates for topological superconductors\cite{fu1, YS}, Cu intercalated Bi$_2$Se$_3$\cite{wray, Hor, Ando, Kriener, Bay}, derived from the topological insulator Bi$_2$Se$_3$ has been investigated most widely. It is believed that the ``glue" for Cooper pairing in this system is facilitated by the strong spin-orbit coupling of Bi$_2$Se$_3$. Within this idea, an electron's spin is locked to the orbital component and the momentum of the electron. Consequently, a short-ranged and spin-independent bare interaction in Cu intercalated Bi$_2$Se$_3$ is spontaneously converted to a spin-dependent and momentum dependent effective interaction between the Bloch electrons thereby leading to superconductivity.\citep{wray}

TCIs are fundamentally different from TIs (e.g. Bi$_2$Te$_3$, Bi$_2$Se$_3$) in terms of the protection mechanism of their surface states. TI surface states are protected by time reversal symmetry, while TCI surface states are protected by crystal symmetry\cite{Fu, Tian, Xu, Hsieh}. It is also known that the TCI surface states are more tunable than that of TIs by perturbations\citep{fu1}. 
Recently, the solid solution composition Pb$_{0.6}$Sn$_{0.4}$Te derived from PbTe ($E_g \sim$ 0.29 eV) and SnTe ($E_g \sim$ 0.18 eV, inverted gap) has created new sensation as a new TCI as confirmed by angle-resolved photoemission spectroscopy (ARPES)\cite{Xu}. Here we show that Pb$_{0.6}$Sn$_{0.4}$Te shows superconductivity in a confined dimension when a mesoscopic point-contact made up of an elemental metal is formed on the surface of Pb$_{0.6}$Sn$_{0.4}$Te. The local superconducting phase shows a high critical temperature $T_c$ that is contact-dependent and varies between 3.7 K and 6.5 K.  Further theoretical studies will be required to understand the microscopic origin of the new phase.

\begin{figure}[h!]
	\centering
		\includegraphics[width=0.7\textwidth]{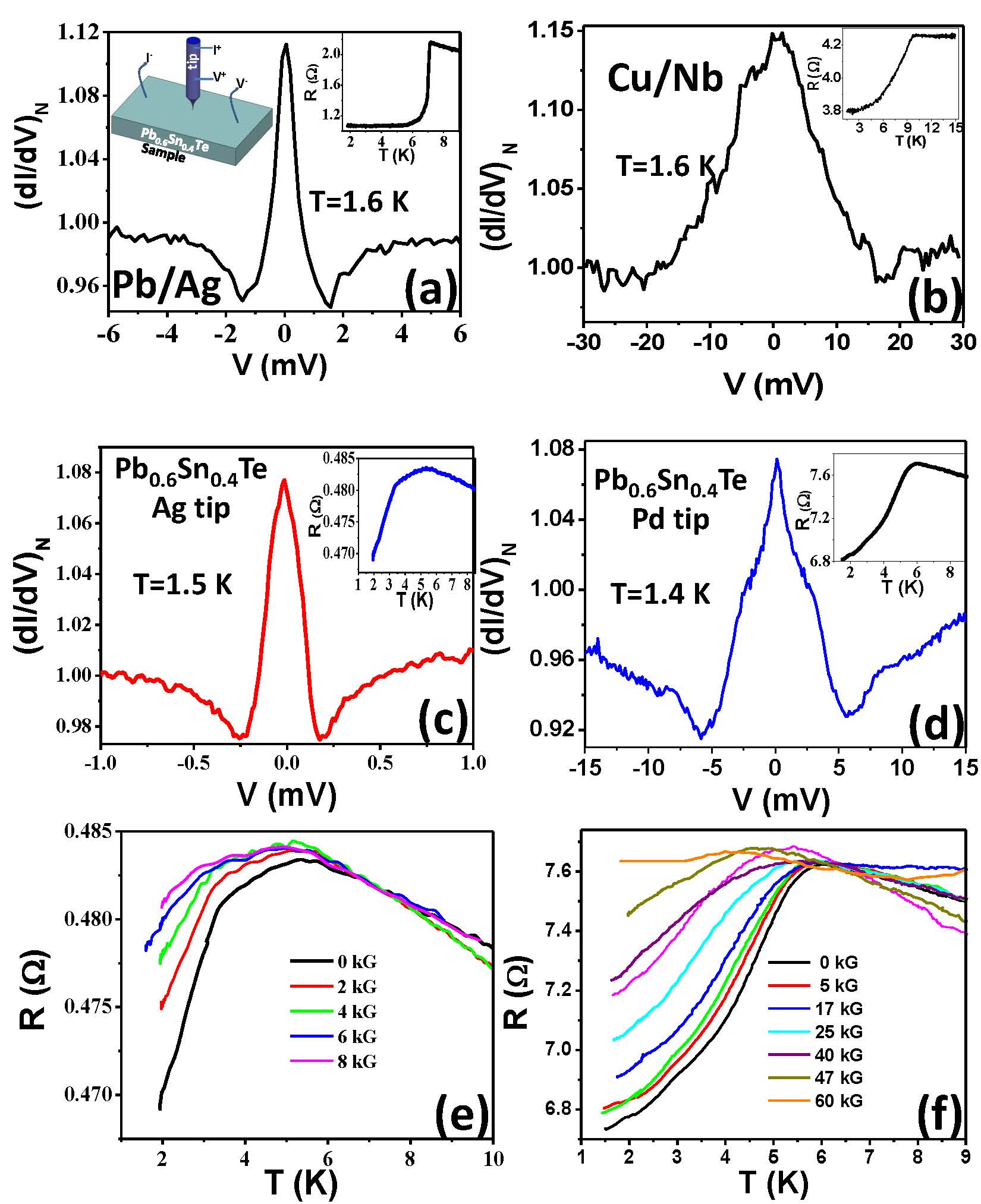}
	\caption{ \textbf{Evidence of superconductivity in Pb$_{0.6}$Sn$_{0.4}$Te point-contacts.} (a) A normalized dI/dV spectrum between Ag and Pb, a conventional superconductor. inset: A schematic diagram showing the point contact on Pb$_{0.6}$Sn$_{0.4}$Te with measuring electrodes (left) and $R-T$ data showing superconducting transition at 7.1 K, the $T_c$ of Pb (right). (b) A normalized dI/dV spectrum between a conventional superconductor Nb and Cu. inset: $R-T$ data showing superconducting transition at 9.4 K, the $T_c$ of Nb (c) A normalized $dI/dV$ spectrum for a point-contact on Pb$_{0.6}$Sn$_{0.4}$Te with an Ag tip. \textit{inset}: $R-T$ data showing superconducting transition at 5.4 K. (d) A normalized $dI/dV$ spectrum for a point-contact between Pb$_{0.6}$Sn$_{0.4}$Te and a Pd tip. \textit{inset}: $R-T$ showing superconducting transition at 6 K. Magnetic field dependence of $R-T$ for point-contacts with (e) an Ag tip and (f) a Pd tip.}
	\label{Figure 1}
\end{figure}

In the inset of Figure 1 (a) we demonstrate a schematic diagram describing how the mesoscopic point contacts between the sharp tips of elemental metals and Pb$_{0.6}$Sn$_{0.4}$Te were formed in-situ at low temperatures. The current and voltage leads are also shown. From the picture it is clear that the measurement is done in a pseudo four-probe geometry. A lock-in based modulation technique was employed to directly measure the differential conductance $dI/dV$ of the point-contacts as a function of an applied DC bias $V$. The resistance $R_{PC}$ of a mesoscopic point-contact between two materials is given by Wexler's formula\cite{wexler}: $R_{PC} = \frac{2h/e^2}{(ak_F)^2} + \Gamma (l/a)\frac{\rho (T)}{2a}$, where $h$ is Planck's constant, $e$ is the charge of a single electron, $a$ is the contact diameter, $\Gamma(l/a)$ is a slowly varying function of the order of unity, $\rho$ is the bulk resistivity of the material and $T$ is the effective temperature at the point-contact. The first term is known as ballistic or Sharvin's resistance ($R_S$) and it is independent of the bulk resistivity of the materials forming the point-contact and independent of temperature. 
The second term is called the Maxwell's resistance ($R_M$) which depends directly on the resistivity of the materials. The above equation suggests when the contact diameter is small compared to electronic mean free path (ballistic regime) $R_S$ dominates and when the contact diameter is large (thermal regime), $R_M$ contributes most to the total contact resistance. Therefore resistive transitions lead to non-linearities in the $I-V$ characteristics corresponding to $R_M$ of point-contacts. Clearly, from Wexler's formula given above, such non-linearities associated with change in $\rho$ are pronounced when $a$ (contact diameter) is large and the point-contacts are in the so-called thermal regime of mesoscopic transport.

One example of a sharp resistive transition is the superconducting transition. This resistive transition can take place when the current flowing through a superconductor exceeds the critical current ($I_c$). If we consider the $I-V$ characteristic of a superconductor, we observe that below $I_c$ the voltage remains zero and above $I_c$ the superconductor slowly becomes resistive thereby making the voltage drop across it finite\cite{tinkham}. When such a $I-V$ curve is differentiated, two sharp dips are expected to appear in the $dI/dV$ vs. $V$ spectra\cite{prb04}. Therefore, two sharp dips symmetric about $V$ = 0 in $dI/dV$ spectra from non-ballistic point-contacts ($a$ large) might indicate the existence of superconductivity at the point-contacts. In addition, since below $I_c$ the heating effect on a thermal regime point-contact is minimized (due to the superconducting transition of one of the electrodes), a sharp increase in zero-bias conductance is also expected\cite{prb04}.

As a test case, in Figure 1(a) and Figure 1(b) we show two $dI/dV$ spectra for point-contacts involving the conventional superconductors lead (Pb) and Nb respectively. Two sharp dips in conductance that are symmetric about $V = 0$ is observed on the Pb/Ag point-contact spectrum due to the critical current of the point-contact. $R-T$ of the point-contact shows a sharp change corresponding to the superconducting transition of Pb at 7.1 K as shown in the inset of Figure 1(a). As it is seen in Figure 1(b), sometimes the critical current driven conductance dips may not appear to be prominent.

The $dI/dV$ spectra obtained on Ag/Pb$_{0.6}$Sn$_{0.4}$Te and Pd/Pb$_{0.6}$Sn$_{0.4}$Te point-contacts are shown in Figure 1(c) and Figure 1(d) respectively. The spectra show two sharp conductance dips symmetric about $V = 0$ as well as a sharp conductance peak at $V = 0$. This is strikingly similar to the data obtained on Pb/Ag point-contacts (Figure 1(b)) where the conductance dips appear due to the critical current of the superconductor and the zero-bias peak appears due to the transition of the superconducting part of the point-contact to a zero-resistance state.\cite{prb04} In addition, as shown in the insets of Figure 1(c) and Figure 1(d), the superconducting transition for the two point-contacts on Pb$_{0.6}$Sn$_{0.4}$Te are clearly seen at temperatures 5.5 K and 6.3 K respectively. We have obtained such results on more than 50 point-contacts on Pb$_{0.6}$Sn$_{0.4}$Te with different elemental metallic tips. Therefore, it is concluded that the metallic point-contacts on Pb$_{0.6}$Sn$_{0.4}$Te are superconducting.

In order to establish the observation of superconductivity on the metallic point-contacts on Pb$_{0.6}$Sn$_{0.4}$Te more firmly it is necessary and sufficient to confirm the validity of the following three key facts about superconducting point-contacts for the point-contacts made on Pb$_{0.6}$Sn$_{0.4}$Te: (i) The superconducting transition as observed in the $R-T$ data shows systematic magnetic field dependence. (ii) The $dI/dV$ spectra show systematic temperature dependence and the spectral features smear out at $T_c$ and (iii) The $dI/dV$ spectra show systematic magnetic field dependence and the spectral features disappear at $H_c$ of a given point-contact. In the following sections we will demonstrate the validity of all the three points mentioned above for the point-contacts made on Pb$_{0.6}$Sn$_{0.4}$Te .

\textbf{(i) $R-T$ data shows systematic magnetic field dependence:} The magnetic field dependence of the $R-T$ data is shown for two point-contacts on Pb$_{0.6}$Sn$_{0.4}$Te, one with a Ag tip (Figure 1(e)) and the other one with a Pd tip (Figure 1(f)). The $R-T$ curves evolve smoothly with magnetic field. As the magnetic field is increased, the transition temperature systematically goes down.

It should be noted here that the over-all resistance of a bulk superconductor can be measured to be zero in a four-probe geometry. However, for a superconducting point-contact\cite{naidyuk}, the contact resistance does not approach zero. This is due to (a) the existence of a non-superconducting component of the point-contact (in this case, a normal metal), (b) existence of a small temperature independent ballistic resistance even when the contacts are far from the ballistic regime and (c) the intrinsic mismatch of the Fermi-velocities in the two electrodes forming the metallic point-contact. Consequently, the existence of superconductivity in a confined geometry under a metallic point-contact cannot be characterized by a directly measurable zero-resistance. This is further clear from the $R-T$ data obtained from point-contacts on superconducting Pb (inset of Figure 1(a)).

\textbf{(ii) The $dI/dV$ spectra show systematic temperature dependence and the spectral features smear out at $T_c$:} In Figure 2(a) we show the temperature dependence of one of the representative $dI/dV$ spectrum obtained on Pd/Pb$_{0.6}$Sn$_{0.4}$Te point-contacts. From the visual inspection alone it is clear that the spectrum smoothly evolves with increasing magnetic field and the spectral features disappear around 6 K, which is the critical temperature as obtained from the transport data at zero magnetic field.

\begin{figure}[h!]
	\centering
		\includegraphics[width=0.6\textwidth]{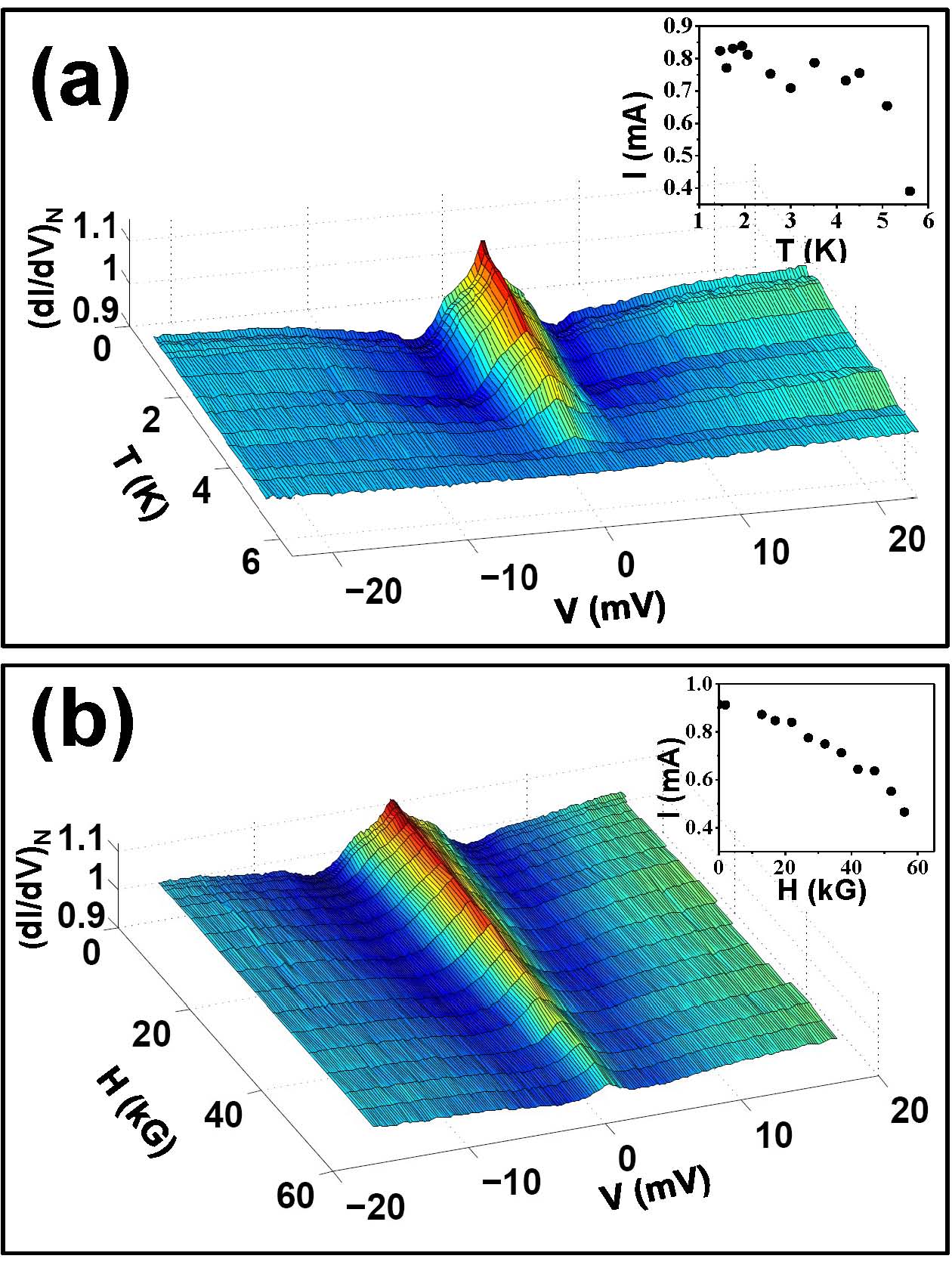}
	\caption{ \textbf{Temperature and magnetic field dependence:} (a) Evolution of the point-contact spectrum between Pb$_{0.6}$Sn$_{0.4}$Te and Pd point-contacts with temperature. \textit{inset:} Temperature dependence of $I_c$. (b) Evolution of the point-contact spectrum with magnetic field. \textit{inset}:  Magnetic field dependence of $I_c$.}
	\label{Figure 2}
\end{figure}

The position of the conductance dips observed in $dI/dV$ gives the estimate of the critical current. This is nothing but the ratio of the magnitude of the DC voltage ($V$) for which the conductance dips appear and the value of the normal state resistance. The critical current thus measured is plotted as a function of temperature in the inset of Figure 2(a). As expected for superconducting point-contacts, the critical current decreases with increasing temperature. Using Wexler's formula and the measured normal state resistance it is also possible to estimate the diameter of the point-contact which is found to be 50 nm in this case. Using this length scale the critical current density is estimated to be of the order of $10^6$ A/cm$^2$ which is reasonable for superconducting point-contacts.

\textbf{(iii) The $dI/dV$ spectra show systematic magnetic field dependence and the spectral features disappear at $H_c$ of a given point-contact:} In order to inspect the validity of this point we show the magnetic field dependence of an Pd/Pb$_{0.6}$Sn$_{0.4}$Te point-contact spectrum in Figure 2(b). As expected for superconducting point-contacts, the spectrum shows a monotonic decrease in the spectral features with increasing magnetic field. Beyond a critical magnetic field of 5 T, the spectrum becomes featureless and that is the critical magnetic field for the superconducting point-contact. The magnetic field dependence of the critical current is demonstrated in the inset of Figure 2(b).

Therefore, we have confirmed the existence of a superconducting phase in Pd/Pb$_{0.6}$Sn$_{0.4}$Te point-contacts.

Having confirmed the existence of a superconducting phase it is now important to investigate the nature of superconductivity in the new phase. In Figure 3(a) and Figure 3(b) we show the $H-T$ phase diagrams constructed for the superconducting phases appearing at Ag/Pb$_{0.6}$Sn$_{0.4}$Te and Pd/Pb$_{0.6}$Sn$_{0.4}$Te point-contacts respectively. In both cases the experimentally obtained $H-T$ curves, within the range of the measurement errors, are seen to fall on the empirically expected $H-T$ phase curves for conventional superconductors.

\begin{figure}[h!]
	\centering
		\includegraphics[width=.7\textwidth]{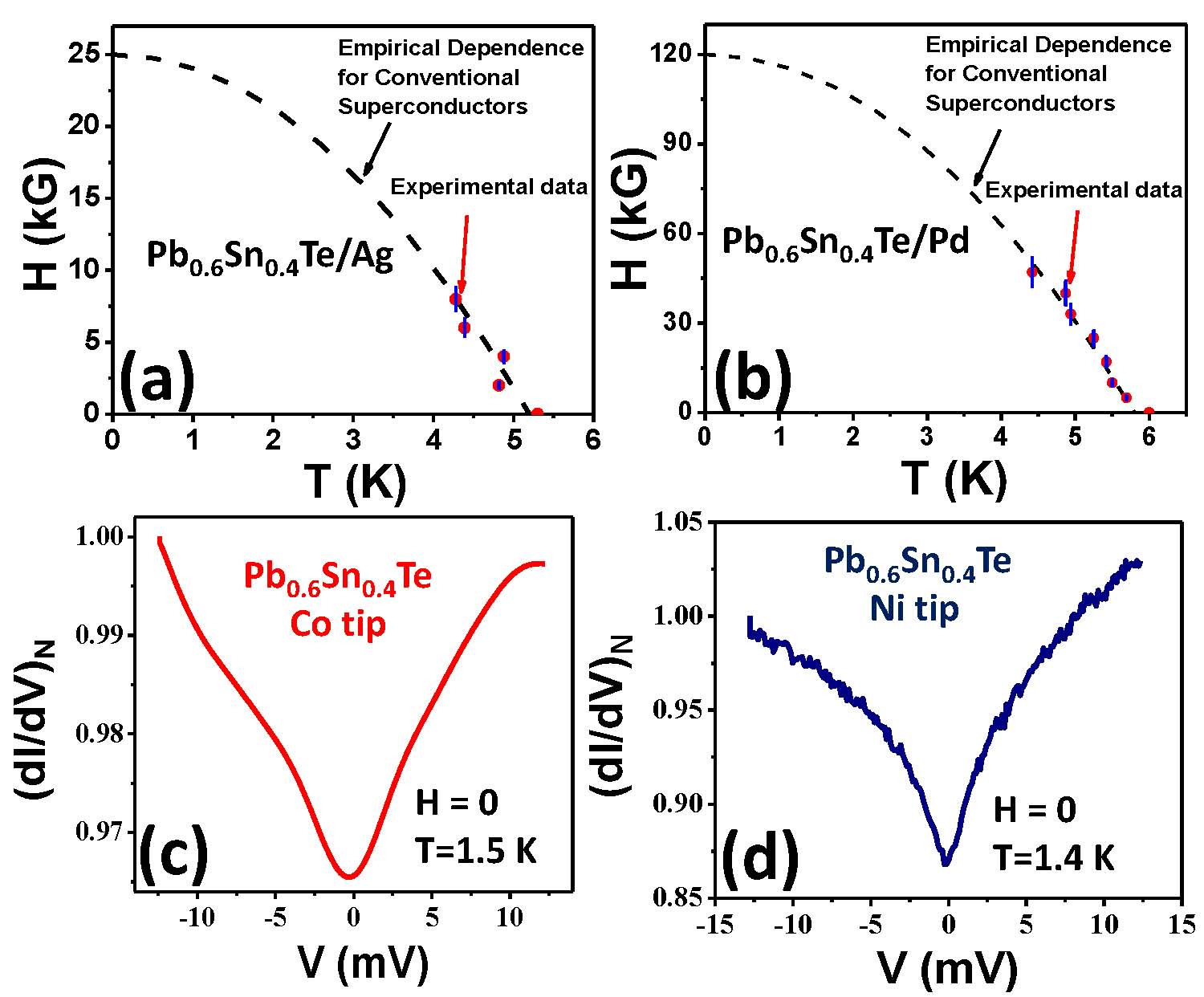}
	\caption{$H-T$ phase diagram obtained for (a) Ag and (b) Pd point-contacts on Pb$_{0.6}$Sn$_{0.4}$Te. These data have been extracted from Figure 1(e) and Figure 1(f) respectively (Error bars are also shown). The dotted lines show the empirical dependence expected for conventional superconductors. Representative point-contact spectrum (c) with a cobalt tip and (d) with a nickel tip on Pb$_{0.6}$Sn$_{0.4}$Te.  }
	\label{Figure 3}
\end{figure}

Since the superconducting phase has been derived from a topologically non-trivial system the possibility of a $p$-wave symmetry of the superconducting order parameter must be addressed too.\cite{majoran1} Since for a $p$-wave symmetry, the time-reversal symmetry is naturally broken, the proximity of a spin-polarized Fermi surface should favour superconductivity. This is in contrast to the conventional BCS-type of superconductors where the proximity of a ferromagnet competes with and suppresses the superconducting order. We report our experimental results on Pb$_{0.6}$Sn$_{0.4}$Te with two metallic ferromagnetic tips Co and Ni in Figure 3(c) and Figure 3(d) respectively. As it is seen, with the ferromagnetic tips, the $dI/dV$ spectrum shows a sharp dip at zero-bias indicating the presence of a gap structure in the density of states. However, no signature of Andreev reflection or critical current dominated effects could be found. At this point the change in the over all shape of the spectra in the proximity of a spin-polarized metal is not understood and this calls for detailed theoretical investigation of spin-polarized transport at mesoscopic interfaces on a topological crystalline insulator.  On the other hand the over all shape of the point-contact spectra presented here and their magnetic field dependence show remarkable similarities with the spectra earlier obtained on Cu-intercalated Bi$_2$Se$_3$.\cite{YS} In that case the sharp zero-bias conductance peak and the unique magnetic field dependence of the same was attributed to the existence of Majorana bound states in Cu-intercalated Bi$_2$Se$_3$.\cite{YS} 

In this context it is also important to discuss other possible origins of a ZBCP other than superconductivity. In the past it was shown that in semiconducting quantum dot systems a ZBCP might originate due to Kondo effect. However, in such cases the peak is expected to show logarithmic temperature dependence and a splitting with magnetic field.\cite{cho,der} Another possibility is a modulation of the density of states at zero-bias due to effects other than superconductivity.\cite{Zeljkovic, druppel, dekker} However, such effects are not expected to show the unique temperature and magnetic field dependence that we observe here. Furthermore, the fact that for certain point contacts we also observe clear signature of Andreev reflection also confirms that the possibilities other than superconductivity can be ruled out for the emergence of the ZBCP in our case.

In principle, the idea of the superconducting order parameter here could be confirmed by measuring the superconducting energy gap through Andreev reflection spectroscopy as a function of temperature with non-ferromagnetic tips. We have attempted to obtain Andreev reflection dominated spectra in the ballistic limit of transport. For such measurements it is required to make the point-contact size smaller than the mean free path. For our Pb$_{0.6}$Sn$_{0.4}$Te samples, the pure ballistic regime could not be achieved possibly due to intrinsic disorder existing in the samples. However, as shown in the supplementary material (Figure S11), we obtained strong signature of Andreev reflection (please note the double peak structure in $dI/dV$ symmetric about $V = 0$) in the last four spectra in Figure S11. These spectra also have strong non-ballistic contribution (confirmed by the sharp dips in $dI/dV$) and consequently could not be used for energy resolved spectroscopy.

Since the topological materials in general have low carrier density the observation of a high critical temperature of 6.5 K is intriguing. As per existing theories the expected transition temperature in such systems is very low and mostly in the milikelvin regime. Furthermore, the existence of a superconducting phase only in a confined geometry with such a high critical temperature on the surface of a topological crystalline insulator is even more intriguing. In this context it may be noted that earlier a superconducting phase was shown to exist at the mesoscopic point-contacts formed on the three dimensional Dirac semi-metal Cd$_3$As$_2$\cite{Cd3As2}.

Stoichiometrically pure Pb$_{0.6}$Sn$_{0.4}$Te samples were used for all the measurements presented here. Since Pb and Sn are also known to superconduct, detailed elemental analysis have been performed to rule out the possibility of obtaining a superconducting signal from local clustering of Pb or Sn.\cite{EPAPS} Moreover, the transition temperature and the critical fields that we measure are significantly different from that expected for Pb or Sn. It should also be noted that the samples used for the presented measurement were polycrystalline in nature. However, as shown in the supplementary material (Figure S5), the grain size of the polycrystals are large ($\sim 100 \mu$). Therefore, majority of the times the point-contact was established on a single crystallite. Since we have obtained the signature of superconductivity at all points on the polycrystalline samples with large crystallites (Please see Figure S11 in the supplementary material), it is reasonable to conclude that all the crystal facets of Pb$_{0.6}$Sn$_{0.4}$Te show superconductivity under metallic point-contacts.

In conclusion, we have provided spectroscopic indication of the emergence of a novel superconducting phase at the mesoscopic point-contacts on the topological crystalline insulator Pb$_{0.6}$Sn$_{0.4}$Te.
The discovery adds an important new candidate in the class of superconductors derived from topologically non-trivial systems.
The results will be extremely helpful in understanding the origin of unexpected superconductivity with relatively high critical temperatures in topological materials in general.\\

MA acknowledges financial support from CSIR, India. GS and KB acknowledge partial financial support from the Ramanujan fellowship, Department of Science and Technology (DST), Govt of India. KB acknowledges support from Sheik Saqr Laboratory and New Chemistry Unit, JNCASR. GS acknowledges financial support from DST Nanomission under the grant number SR/NM/NS-1249/2013.\\


\beginsupplement

\newpage
\textbf{Supplemental material for "Unexpected superconductivity at nanoscale junctions made on the topological crystalline insulator Pb$_{0.6}$Sn$_{0.4}$Te"}\\\\

(I) \textbf{Reagents:} Tin (Sn, Alfa Aesar 99.99+ $\%$), tellurium (Te, Alfa Aesar 99.999+ $\%$) and lead (Pb, Alfa Aesar 99.99+ $\%$) were used for synthesis without further purification.\\

(II) \underline{Material synthesis:} High quality ingots ($\sim$1 g) of Pb$_{0.6}$Sn$_{0.4}$Te was synthesized by mixing appropriate ratios of high-purity elemental Sn, Pb and Te in quartz tube. The tube was  sealed under vacuum ($10^{-5}$ Torr) and slowly heated to 723 K over 12 hrs, then heated up to 1323 K in 5 hrs, soaked for 5 hrs, and cooled down to 1023 K over 2 hrs and soaked for 4 hrs, then slowly cool down to room temperature over a period of 18 hrs.\\ 

(III) \underline{Structural and chemical analysis of materials:}\\

\textbf{Powder X-ray diffraction:} Powder X-ray diffraction for finely grinded sample was recorded
using a Cu K$_{\alpha}$(λ = 1.5406 $A^{o}$) radiation on a Bruker D8 diffractometer. Powder X-ray diffraction (PXRD) pattern of as synthesized polycrystalline $Pb_{0.6}Sn_{0.4}Te$ could be index based on cubic PbTe structure (space group, Fm-3m). PXRD of  $Pb_{0.6}Sn_{0.4}Te$ are in well match with simulated PXRD pattern from the single crystal data of closest composition $Pb_{0.7}Sn_{0.3}Te$ (Figure S1a, supporting information). We have further compared  PXRD data of $Pb_{0.6}Sn_{0.4}Te$ with pristine $PbTe$ and $SnTe$ in Figure S1b and 1c in supporting information. Highest intensity (200) peak of $Pb_{0.6}Sn_{0.4}Te$  lies in between that of $PbTe$ and $SnTe$ (Figure 1c in supporting information), which confirms complete solid solution formation $Pb_{0.6}Sn_{0.4}Te$, thus no phase separation occurs.
\newpage

\begin{figure}[h!]

	\centering
		\includegraphics[width=.7\textwidth]{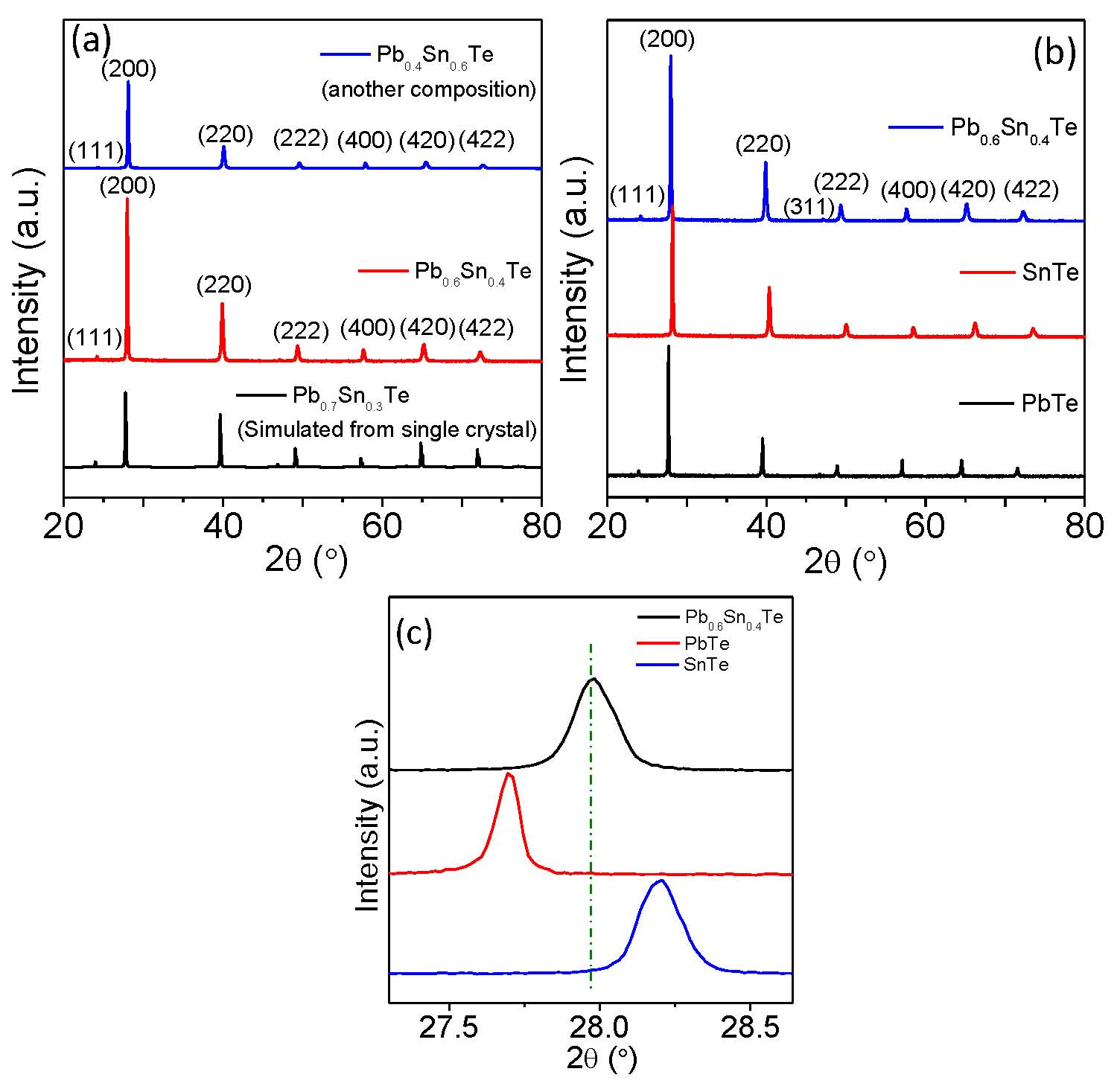}
	\caption{(a) Powder X-ray diffraction patterns of $Pb_{0.6}Sn_{0.4}Te$ and $Pb_{0.4}Sn_{0.6}Te$ samples. Simulated PXRD from the single crystal of $Pb_{0.7}Sn_{0.3}Te$ are given for comparison. (b) Powder X-ray diffraction patterns of $Pb_{0.6}Sn_{0.4}Te$, $PbTe$ and $SnTe$. (c) Zoomed XRD patterns from 27.2 to 28.6 degree for Figure (b).  }
	\label{Figure 3}
\end{figure}

\begin{figure}[h!]
	\centering
		\includegraphics[width=.5\textwidth]{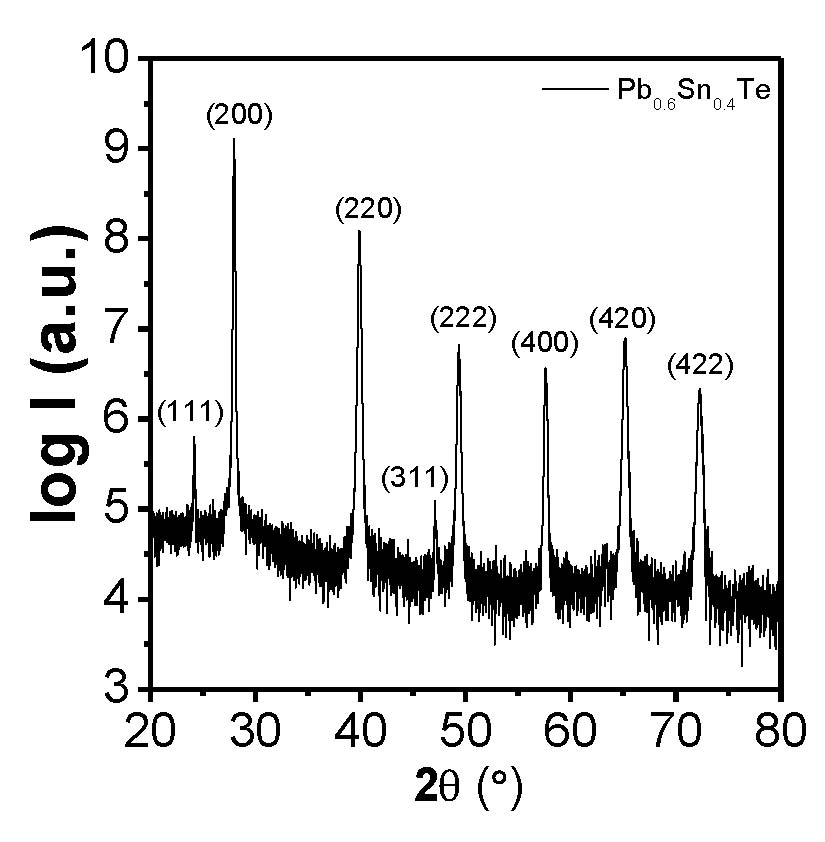}
	\caption{(a) Powder X-ray diffraction patterns of $Pb_{0.6}Sn_{0.4}Te$ in logarithmic scale.}
	\label{Figure 3}
\end{figure}

\newpage

\textbf{X-ray photoelectron spectroscopy:} XPS measurement has been performed with $Mg K_{\alpha}(1253.6 eV)$ X-ray source with a relative composition detection better than 0.1$\%$ on an Omicron Nano-technology spectrometer. The coordination number of Te is six and Te is bonded to $Sn$ as well as $Pb$ in $Pb_{0.6}Sn_{0.4}Te$. Thus, $Te$ experiences different chemical environment from $Pb$ and $Sn$. Difference of electronegativity between $Sn$ and $Te$ is small compared to that of $Pb$ and $Te$. Covalent nature of $SnTe$ is higher than $PbTe$. Thus, $Te$ bonded with $Sn$ has higher binding energy than that of bonded with $Pb$. Therefore, $Te$ 3d5/2 and 3d3/2 peaks are seen to be spitted in two peaks.\\

\begin{figure}[h!]
	\centering
		\includegraphics[width=\textwidth]{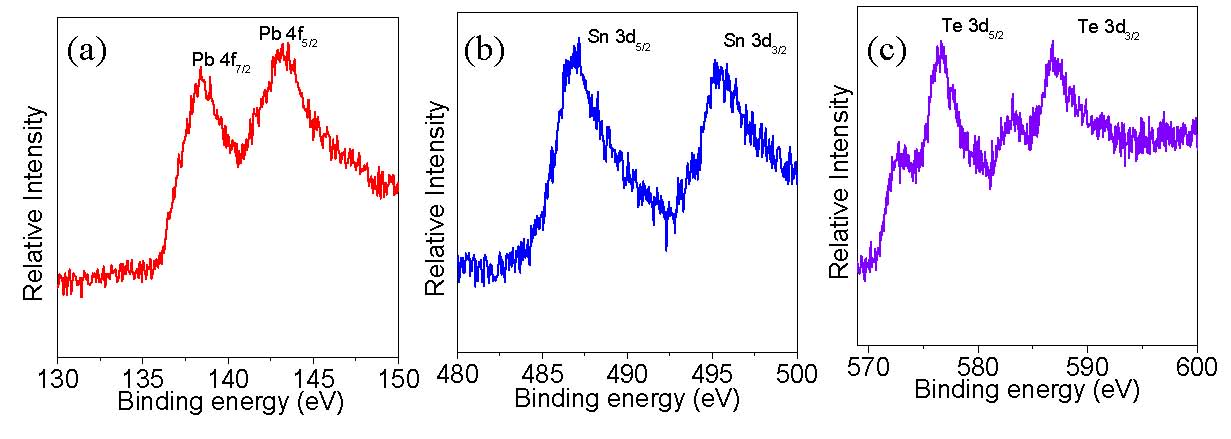}
	\caption{XPS spectra of $Pb_{0.6}Sn_{0.4}Te$. (a) Pb (2+) 4f, (b) Sn (2+) 3d, and (c) Te (2-) 3d spectra. }
	\label{Figure 3}

\end{figure}
\newpage
\textbf{Energy Dispersive X-ray analysis:} EDAX  compositional analysis was performed during FESEM imaging. Errors  in  the  determination  in  compositions  of Pb$_{0.6}$Sn$_{0.4}$Te in EDAX measurements is nearly 5 $ \%$.
\begin{figure}[h!]
	\centering
		\includegraphics[width=.7\textwidth]{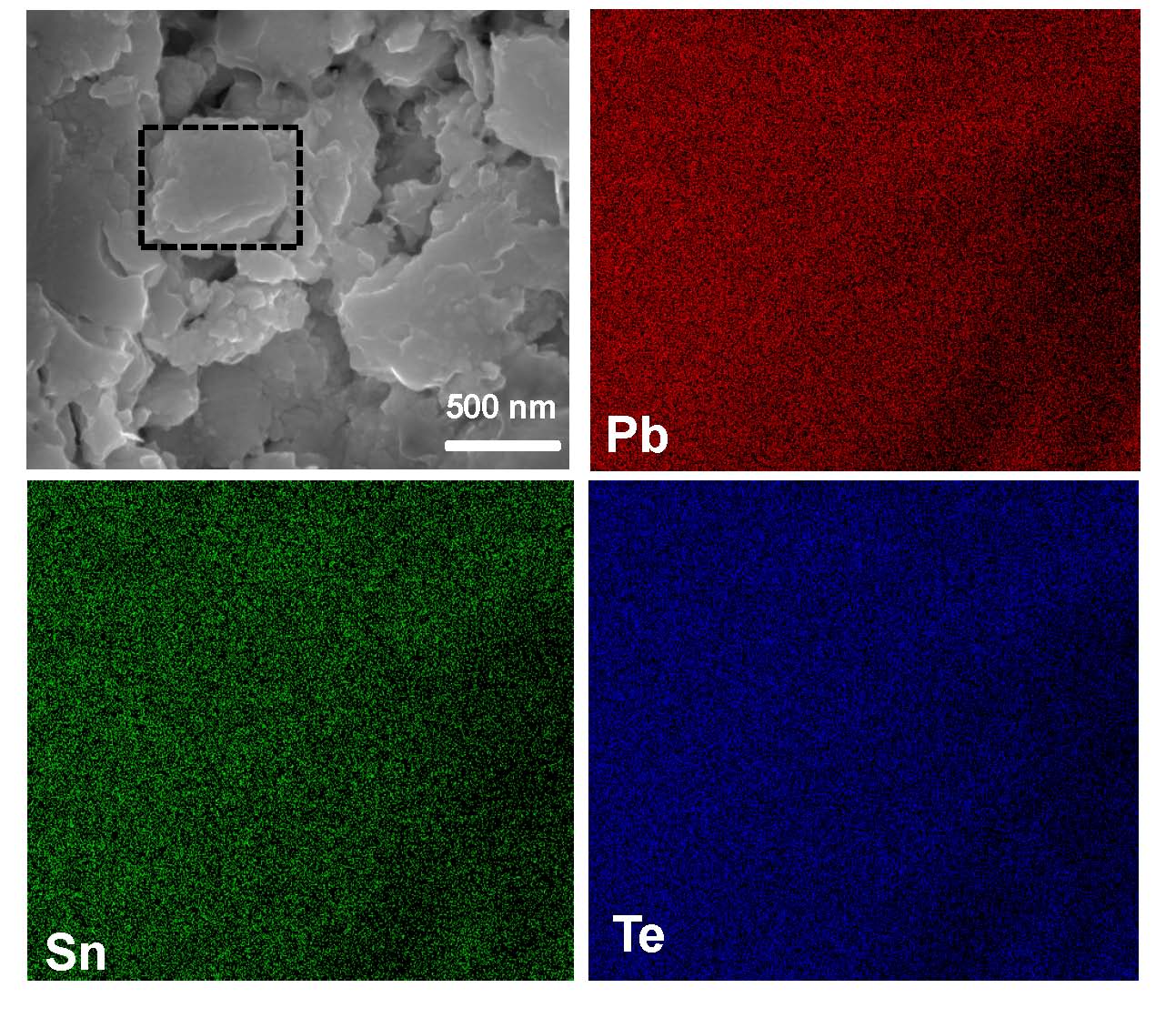}
	\caption{(a) EDAX color mapping of freshly cleaved  surface of $Pb_{0.6}Sn_{0.4}Te$  showing  the presence of all elements (Pb, Sn, Te). }
	\label{Figure 3}
\end{figure}
\newpage
\textbf{Field emission scanning electron microscopy:} FESEM experiments were performed using NOVA NANO SEM 600 (FEI, Germany) operated at 15 KV. $Pb_{0.6}Sn_{0.4}Te$ exhibits the grain size more that 100 $\mu$m as can be observed from the field emission scanning electron microscopy image (Figure S5, Supplementary material) measured under back-scattering mode.

\begin{figure}[h!]
	\centering
		\includegraphics[width=.5\textwidth]{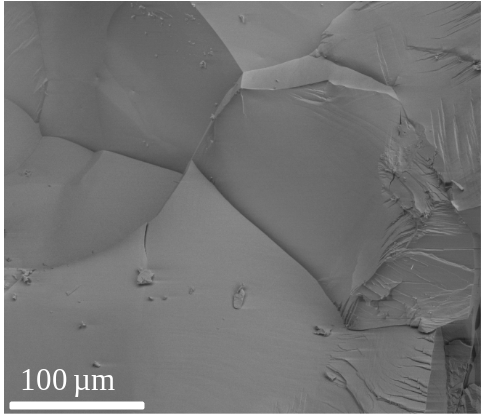}
	\caption{ Backscattered mode FESEM image of Pb$_{0.6}$Sn$_{0.4}$Te sample showing the  grain size bigger than 100 $\mu$m. }
	\label{Figure 3}

\end{figure}

(IV) \underline{Low-temperature measurements:} The point-contact spectroscopy measurements were performed in a liquid helium cryostat equipped with a dynamic and a static variabale temperature insert (VTI) that works in a temperature range between 1.4 K and 300 K. The same cryostat is also equipped with a three axis vector magnet where the vertical field (perpendicular to the sample holder disc) is generated by a superconducting solenoid. All the in-field measurements presented in this paper were perfomred by charging the solenoid up to a field of 6 T. 

Electrodes of 50 $\mu$m gold wire were mounted on Pb$_{0.6}$Sn$_{0.4}$Te  with silver epoxy (2 component, epotek) and mounted on a disc-shaped sample holder made up of copper. A cernox thermometer and a 50 $\ohm$ cartridge heater were directly mounted on the same copper piece for controlling the sample temperature precisely. A Lakeshore temperature controller (Model no. 350) was used to stabilize the temperature with predetermined P, I, D parameters. The point-contacts between the normal metals and the sample were formed by approaching a sharp metallic tip using a coarse approach mechanism based on a 100 threads per inch (t.p.i.) differential screw purchased from Thorlab. 

The $dI/dV$ measurements were done using a SRS 830 lock-in amplifier. The output of the lock-in amplifier was converted to an ac excitation current ($i_{ac}cos \omega t$) that was electronically coupled to a DC current ($I_{dc}$) generated from a Keithely (model no. 6221) current sourse. The total current $I = I_{dc} + i_{ac}cos \omega t$ was sent through the point-contact. The voltage drop across the point-contact was measured in a pseudo four-probe geometry using a digital multimeter (Keithley: 2000) for the dc component ($V$) and the same lock-in amplifier as mentioned above (locked at a the frequency $\omega$) for the ac component ($v_{ac}$). The magnitude of $v_{ac}$ was kept in the range between $10 \mu V$ and $50 \mu V$. In this limit $v_{ac}$ can be approximated to be proportional to differential change in voltage $dV$ for a given $V$. The differential change in current $dI$ is measured directly for plotting $dI/dV$ spectra. This technique is traditionally known as lock-in modulation technique for direct measurement of differential conductance.\\

(V) \underline{Four probe resistivity :}
\begin{figure}[h!]
	\centering
		\includegraphics[width=.5\textwidth]{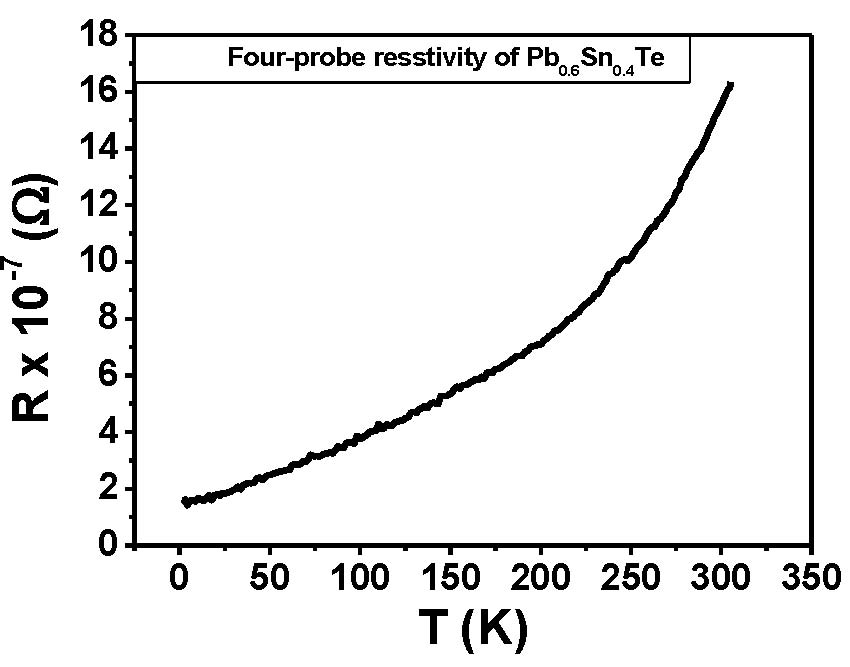}
	\caption{Four-probe resistance vs. Temperature of  of $Pb_{0.6}Sn_{0.4}Te$ }
	\label{Figure 3}
\end{figure}\newpage

(VI) \underline{Spectroscopic measurement :} 
\begin{figure}[h!]
	\centering
		\includegraphics[width=.5\textwidth]{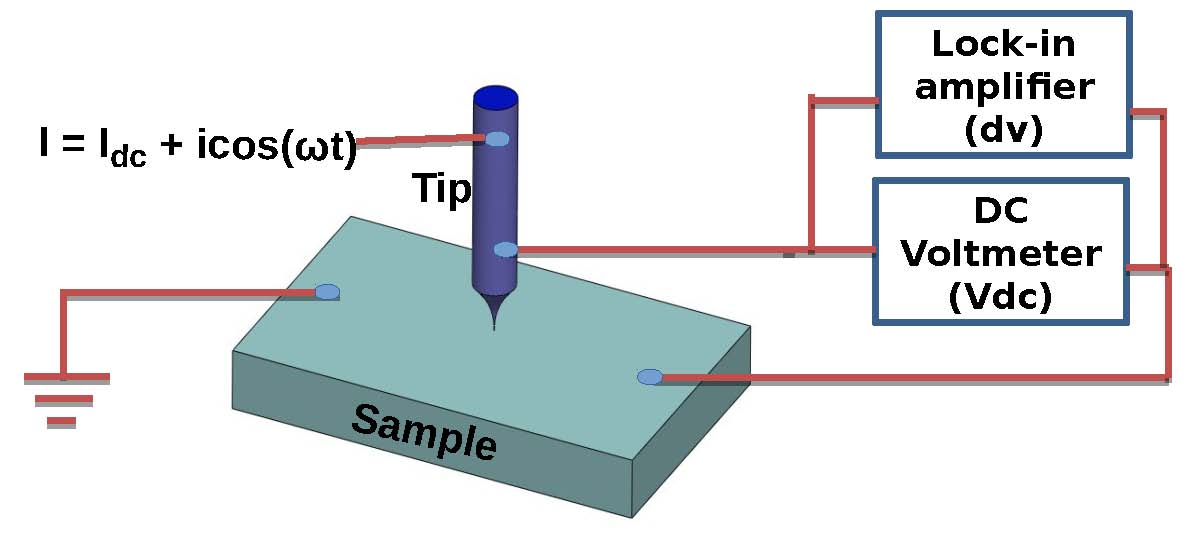}
	\caption{Schematic diagram of point-contact spectroscopy measurement }
	\label{Figure 3}
\end{figure}

(VII) \underline{Additional point-contact data with temperature and magnetic field evolution:}
\begin{figure}[h!]
	\centering
		\includegraphics[width=.4\textwidth]{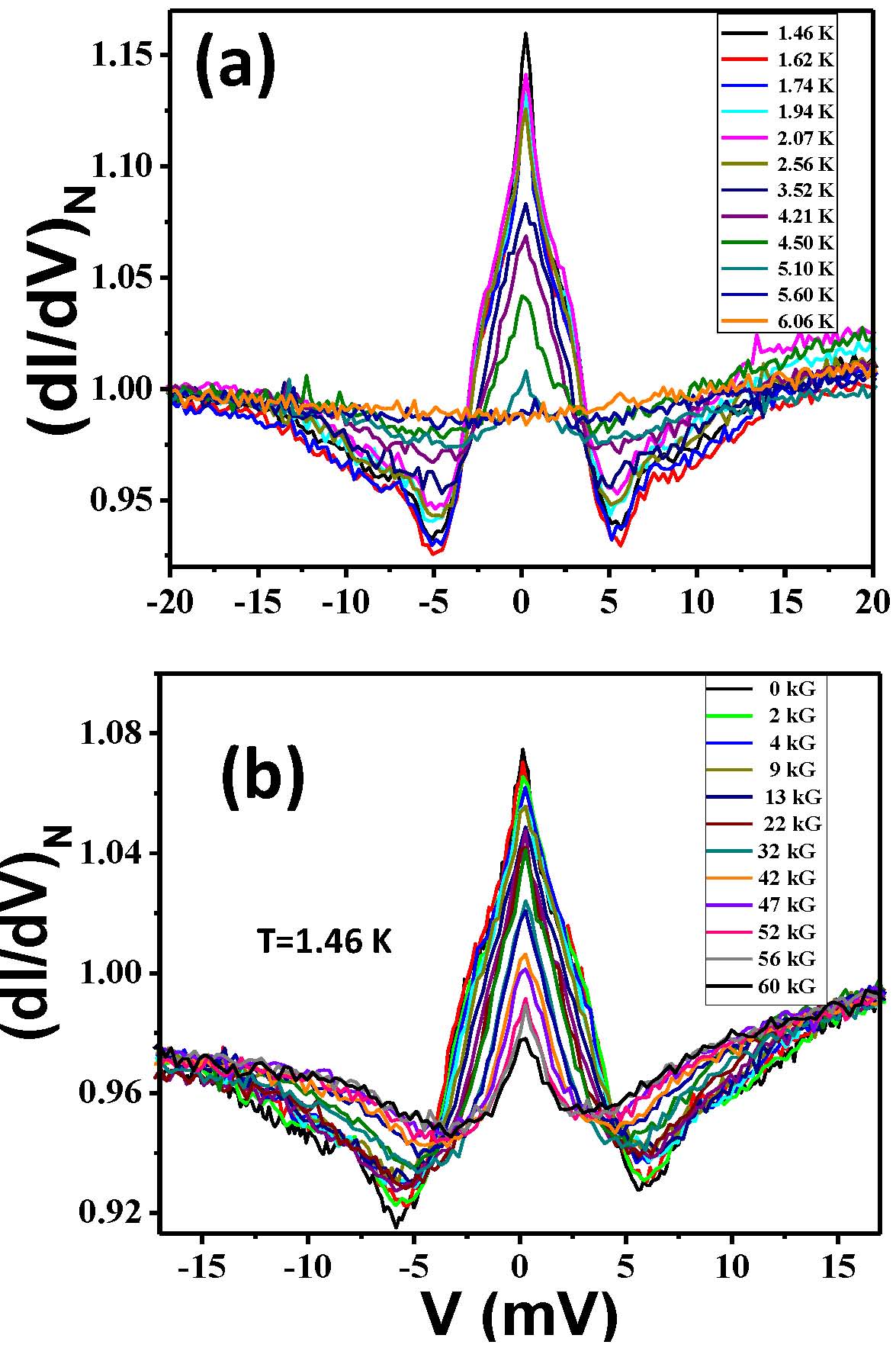}
	\caption{  \textbf{Temperature and magnetic field dependence between Pb$_{0.6}$Sn$_{0.4}$Te and Pd point-contacts:} Evolution of the point-contact spectrum (a) with temperature (2-D plot corresponding to the figure 2(a) of main manuscript) (b) with magnetic field (2-D plot corresponding to the figure 2(b) of main manuscript) }
	\label{Figure 3}
\end{figure}

\begin{figure}[h!]
	\centering
		\includegraphics[width=.5\textwidth]{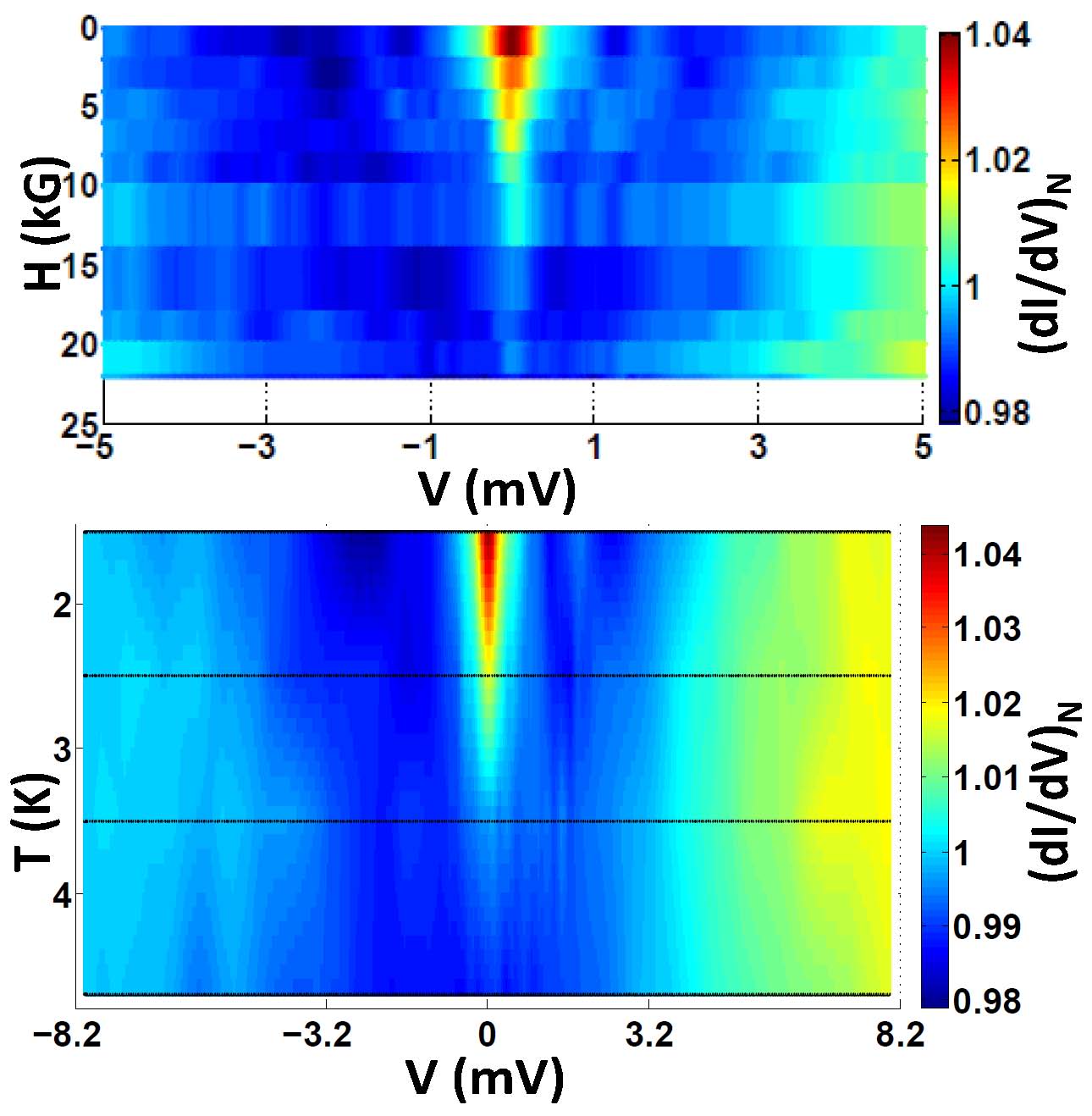}
	\caption{ \textbf{Temperature and magnetic field dependence between Pb$_{0.6}$Sn$_{0.4}$Te and Ag point-contacts:} Evolution of the point-contact spectrum (a) with magnetic field (b) with temperature.}
	\label{Figure 3}
\end{figure}\newpage

(VIII) \underline{Statistical evaluation of $T_c$ : }

\begin{table}[h!]
	\centering
		\includegraphics[width=.6\textwidth]{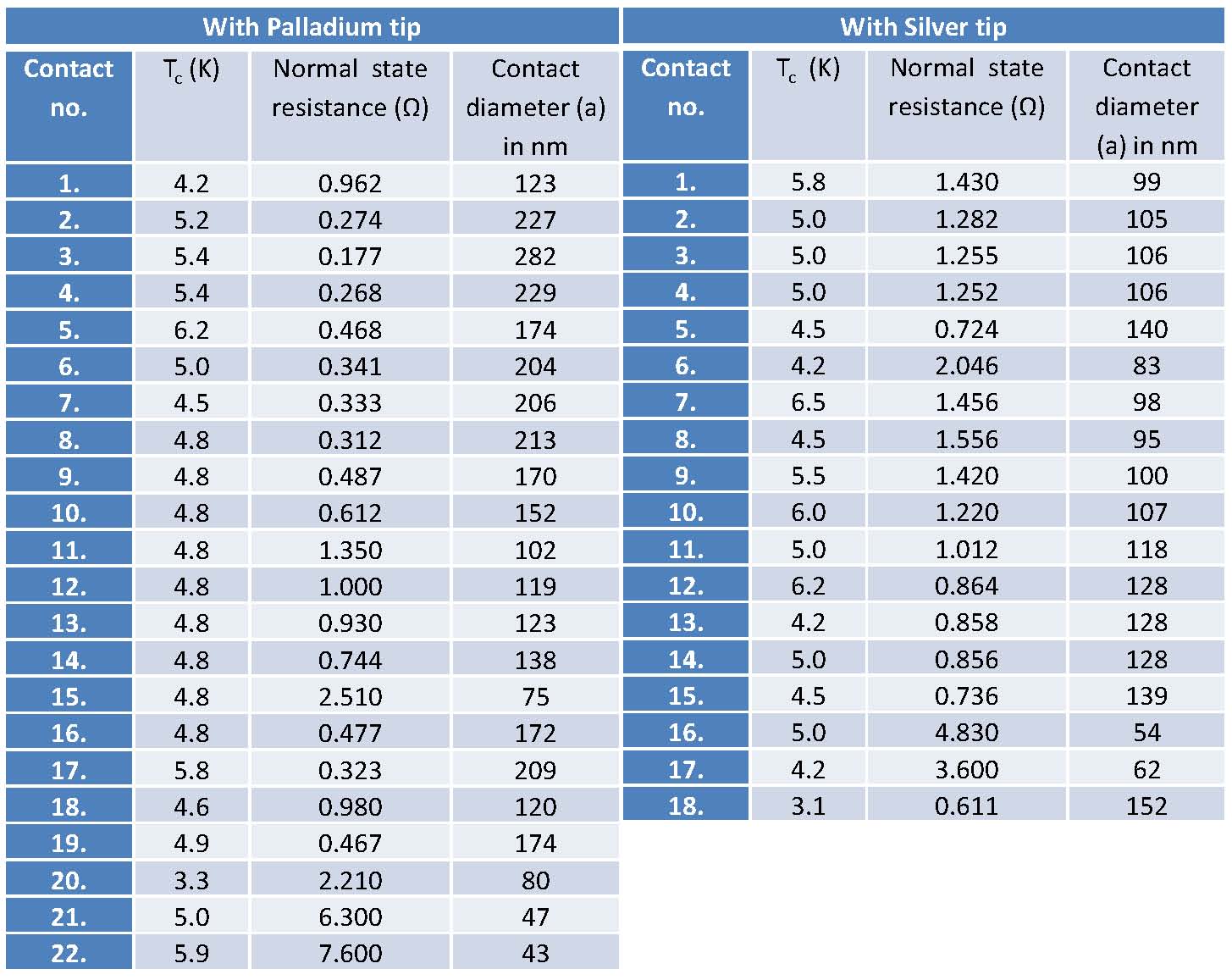}
	\caption{$T_c$ of different point-contacts with Pd (left)and Ag (right) tips  }
	\label{Table}
\end{table}

\begin{figure}[h!]
	\centering
		\includegraphics[width=.5\textwidth]{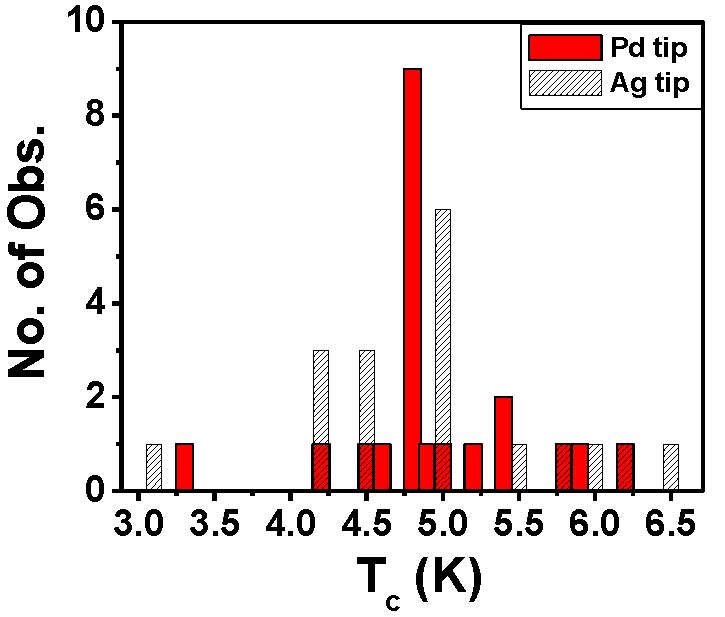}
	\caption{Bar diagram demonstrating  $T_c$ at different point-contacts with Pd (red in colour) and Ag (shaded) tips }
	\label{Figure 3}
\end{figure}\newpage

(VIII) \underline{Representative spectra obtained at different points on the sample: }

\textbf{(Please see the next page)}

\begin{figure}[h!]
	\centering
		\includegraphics[width=.95\textwidth]{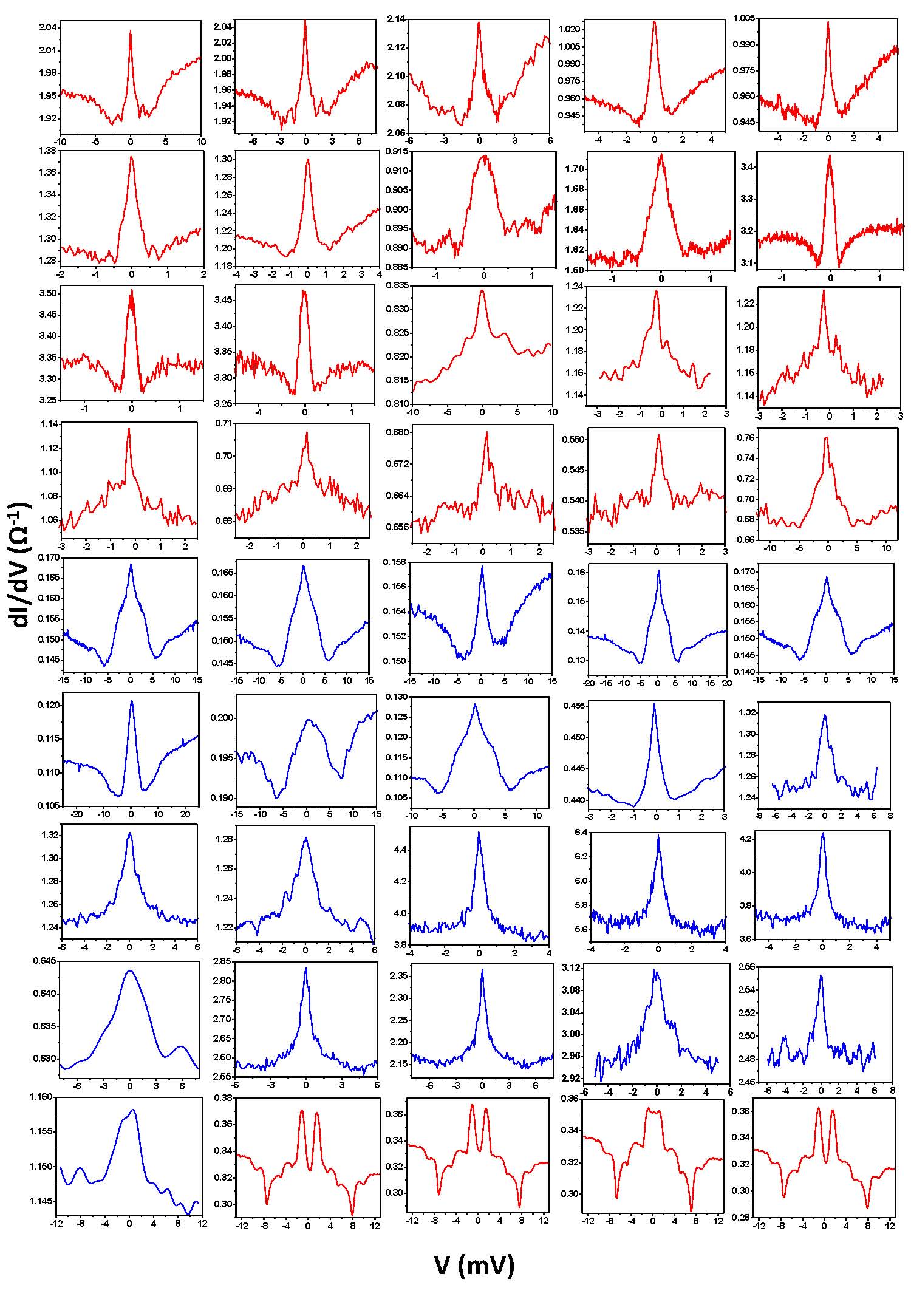}
	\caption{dI/dV vs Vdc plots at different point-contacts with Ag (red in color) and Pd (blue in color) tips }
	\label{Figure 3}
\end{figure}



\begin{thebibliography}{100}




\bibitem{Peng} Peng H. \textit{et al}., \textit{Phys. Rev. B.}, \textbf{88} (2013) 024515.



\bibitem{Kirzhner} Kirzhner T., Lahoud E., Chaska K. B., Salman Z. and  Kanigel A., \textit{Phys. Rev. B.,} \textbf{86} 2012 064517.



\bibitem{Satoshi} Sasaki S. \textit{et al}., \textit{Phys. Rev. Lett.}, \textbf{107} (2011) 217001.



\bibitem{YS} Hor Y. S., \textit{et al}., \textit{Phys. Rev. Lett.}, \textbf{104} (2010) 057001.



\bibitem{rmp1} Hasan M. Z. and Kane C. L., \textit{Rev. Mod. Phys.}, \textbf{82} (2010) 3045.




\bibitem{rmp2} Qi Xiao-Liang and Zhang Shon-Cheng, \textit{Rev. Mod. Phys.}, \textbf{83} (2011) 1057.




\bibitem{sato} Sato T. \textit{et al}., \textit{Phys. Rev. Lett.}, 110 (2013) 206804.




\bibitem{fu1} Ando Y. and Fu L., \textit{Cond. Mat. Phys.}, \textbf{6} (2015) 361.





\bibitem{Fu} Fu L., \textit{Phys. Rev. Lett.}, \textbf{106} (2011) 106802.




\bibitem{liu1} Liu J. \textit{et al}., \textit{Nat. Mater.}, \textbf{13} (2014) 178.




\bibitem{Zeljkovic} Zeljkovic I. \textit{et al}., \textit{Nature Materials}, \textbf{14} (2015) 318-324.




\bibitem{Tian} Liang T. \textit{et al}., \textit{Nature Communications}, \textbf{4} (2013) 2696.




\bibitem{Mitrofanov} Mitrofanov K. V. \textit{et al}., \textit{J. Phys.: Condens. Matter.}, \textbf{26} (2014) 475502.



\bibitem{Ilija} Zeljkovic I. \textit{et al}., \textit{Nature Physics}, \textbf{10} (2014) 572-577.




\bibitem{okada} Okada Y. \textit{et al}., \textit{Science}, \textbf{341} (2013) 1496.



\bibitem{Erickson} Erickson A. S., Chu J. H. M.,  Toney F., Geballe T. H. and Fisher I. R., \textit{Phys. Rev. B.}, \textbf{79} (2009) 024520.



\bibitem{balu} Balakrishnan G., Bawden L., Cavendish S. and Lees M. R., \textit{Phys. Rev. B.}, \textbf{87} (2013) 140507(R).




\bibitem{Leijnse} Leijnse Martin, Flensberg Karsten, \textit{Semicond. Sci. Technol.}, \textbf{27} (2012) 124003.




\bibitem{majoran1} Fu L. and Kane C. L., \textit{Phys. Rev. Lett.}, \textbf{100} (2008) 096407.




\bibitem{Been} Beenakker C. W. J., \textit{Annu. Rev. Con. Mat. Phys.}, \textbf{4} (2013) 113-136.




\bibitem{wray} Wray L. Andrew, \textit{et al.}, \textit{Nature Physics,} \textbf{6} (2010) 855-859.




\bibitem{Hor} Hor Y. S., Checkelsky J. G., Qu D., Ong N. P. and Cava R. J., \textit{J. Phys. Chem. Solids}, \textbf{72} (2011) 572-576.



\bibitem{Ando} Kriener M., Segawa Kouji, Ren Zhi, Sasaki Satoshi, Wada  Shohei, Kuwabata  Susumu and Ando Yoichi, \textit{Phys. Rev. B.}, \textbf{84} (2011) 054513.




\bibitem{Kriener} Kriener M, Segawa Kouji, Ren Zhi, Sasaki Satoshi and Ando Yoichi, \textit{Phys. Rev. Lett.},  \textbf{106} (2011) 127004.




\bibitem{Bay}Bay T. V., Naka T., Huang Y. K., Luigjes H., Golden M. S. and Visser de A., \textit{Phys. Rev. Lett.}, \textbf{108} (2012) 057001.





\bibitem{Xu} Xu Su-Yang \textit{et al}., \textit{Nature Communications,} \textbf{3} (2012) 1192.




\bibitem{Hsieh} Hsieh h. T., Lin H., Liu J., Duan W., Bansil A. and Fu L., \textit{Nature Communications,}  \textbf{3} (2012) 982.




\bibitem{wexler} Wexler A., \textit{Proc. Phys. Sov.}, \textbf{89} (1966) 927.




\bibitem{tinkham} Tinkham M., Introduction to Superconductivity, Dover, Second Edition (2004).




\bibitem{prb04} Sheet  Goutam, Mukhopadhyay S. and Raychaudhuri P., \textit{Phys. Rev. B.}, \textbf{69} (2004) 134507.




\bibitem{naidyuk} Naidyuk Y. G. and Yanson K. I., Point-Contact Spectroscopy, Springer (2005).


\bibitem{cho} Cho Sungjae, Zhong Ruidan, Schneeloch John A., Gu Genda, and Mason Nadya, \textit{Sci. Rep.,} \textbf{6} (2016) 21767.



\bibitem{der} Wiel W. G. van der, \textit{et al.}, \textit{Science,} \textbf{289} (2000) 2105


\bibitem{druppel} Druppel Matthias, Kruger Peter and Rohlfing Michael, \textit{Phys. Rev. B.,} \textbf{90} 2014 155312.




\bibitem{dekker} Venema L. C., Janssen J. W., Buitelaar M. R., Wildoer J. W. G., Lemay S. G., Kouwenhoven L. P. and Dekker C., \textit{Phys. Rev. B.,} \textbf{62} 2000 5238.



\bibitem{Cd3As2} Aggarwal Leena, Gaurav Abhishek, Thakur Gohil S., Haque Zeba, Ganguli Ashok K., Sheet Goutam,  \textit{Nature Materials.} \textbf{15} (2016) 32-37.





\bibitem{EPAPS} See online supplementary materials.\\\\

\end{thebibliography}
\end{document}